\documentclass[12pt]{article}
\usepackage{amssymb,amsmath,amsthm,amscd,latexsym}
\usepackage{mathrsfs}
\usepackage{mathrsfs}
\usepackage{amsfonts}
\usepackage{amsmath}
\usepackage{amssymb}
\usepackage{amscd}
\usepackage{mathrsfs}
\usepackage{amssymb}
\usepackage{amsmath}
\usepackage{amsthm}
\usepackage{latexsym}
\usepackage{indentfirst}
\usepackage{enumitem}
\usepackage{anysize}
\usepackage{bbm}

\renewcommand{\paragraph}{\roman{paragraph}}
\input cyracc.def

\theoremstyle{definition}

\begin{document}
\title{\bf Pisano period codes
\thanks{This research is supported by National Natural Science Foundation of China (61672036),
Technology Foundation for Selected Overseas Chinese Scholar, Ministry of Personnel of China (05015133) and
the Open Research Fund of National Mobile Communications Research Laboratory, Southeast University (2015D11) and
Key projects of support program for outstanding young talents in Colleges and Universities (gxyqZD2016008).}
}
\author{
\small{Minjia Shi$^{1,2,3}$, Zhongyi Zhang$^4$, and Patrick Sol\'e$^5$}\\ 
\and
\small{${}^1$Key Laboratory of Intelligent Computing \& Signal Processing,}\\
 \small{Ministry of Education, Anhui University No. 3 Feixi Road,}\\
  \small{Hefei Anhui Province 230039, P. R. China;}\\
\small{${}^2$School of Mathematical Sciences, Anhui University, Hefei, 230601, P. R. China}\\  
  \small {${}^3$National Mobile Communications Research Laboratory}\\
\small {Southeast University, 210096, Nanjing, P. R. China;}\\
[-0.8ex]
\small{${}^4$ School of Wendian, Anhui University, Hefei, Anhui, 230601, P. R. China}\\
\small{${}^5$CNRS/LAGA, University of Paris 8, 2 rue de la Libert\'e, 93 526 Saint-Denis, France}
}
\date{}
\maketitle
{\bf Abstract:} {The cyclic codes with parity check polynomial the reciprocal of the characteristic polynomial of the Fibonacci recurrence over a prime finite field are shown to have either one weight or two weights. When these codes are irreducible cyclic
we obtain many counterexamples to the conjectural classification of two-weight irreducible cyclic codes of Schmidt and White (2002). When they are reducible and projective their duals are uniformly packed.
}


\section{Introduction}
The topic of two-weight codes and their many connections with strongly regular graphs, association schemes \cite{BH}, and finite geometries \cite{Cald}, has been explored since the seminal work of Delsarte \cite{D}. To construct two-weight codes an old technique is to use irreducible cyclic codes, which, in turn can be studied by Gauss sums, Fourier transform and L-series \cite{M}. Recently, a classification scheme of two-weight irreducible cyclic codes was announced in \cite{SW}.

In the present paper, we present many counterexamples to that scheme by using the correspondence between linear recurrences and cyclic codes and the oldest known linear recurrence: the Fibonacci sequence first introduced in the XIIIth century by Leonardo da Pisa. The periodic behavior modulo an integer was first observed by Lagrange in the XVIII th century, and studied by Edouard Lucas in the XIXth century \cite{L}. For details and perspective we refer to \cite{R}. We distinguish two cases, depending on the discriminant of the characteristic polynomial $x^2-x-1$ being a square or not in  $\mathbb{F}_{p}.$ In the first case we construct irreducible cyclic codes that have at most two weights. In the second case we construct reducible cyclic codes that are always two-weight codes, and, in some cases, are projective codes with uniformly packed duals. Our methods are elementary and only rely on the finite analogue of the Binet formula. There is another class of dimension 2, two-weight irreducible cyclic codes in \cite{RP}, that contradicts the Schmidt-White scheme, but the proofs there rely on cyclotomic numbers and the examples are restricted to $p\equiv 3 \pmod{4}.$

The material is organised as follows. The next section contains the necessary background to understand Section 3 where the main results are exposed. In that section we study first the case when the check polynomial of our code is irreducible, then in a second subsection the case when it is irreducible. In a third subsection the dual distance is determined  and the spectrum of the coset graph is explored. Section 4 contains numerical examples. The last section concludes the paper.
\section{Background material}
\subsection{Linear recurrences over the integers}
The {\bf Fibonacci numbers} $F_n,$ are defined by the recurrence $F_{n+2}=F_{n+1}+F_n,$ and initial conditions $F_0=0,\,F_1=1.$
The generalised Fibonacci numbers $G_n$ follow the same recurrence with arbitrary initial conditions.
 The {\bf Pisano period} $\pi(m)$ (denoted by $k(m)$ in \cite{R}) attached to an integer $m$ is the period of the Fibonacci sequence modulo $m.$ In general, the period of $G_n$ modulo $m$ divides $\pi(m)$ (\cite[Theorem 3.15]{R}). The sequence $\pi(m)$ is sequence $A001175$ in the database {\tt www.oeis.org}. The {\bf rank} of $F_n$ mod $p$ denoted by $\alpha(p)$ is the least integer $r>0$ such that $p \vert F_r.$ It is easy to see that this number is the multiplicative order of $r/s$ in $\mathbb{F}_{p}.$ The {\bf order} of $F_n$ mod $p$ denoted by $\beta(p)$ is then defined as $\frac{\pi(p)}{\alpha(p)}.$ A striking and nontrivial property of the order is that it can only take the three values $1,\,2,\,4$ \cite[Cor. 2.39]{R}. In the following, to simplify notations we let $N=\pi(p),\,e=\alpha(p),$ and $K=\beta(p).$
 \subsection{Cyclic codes}
 A cyclic code of length $N$ over a finite field $\mathbb{F}_{p}$ is a $\mathbb{F}_{p}$ linear code of length $N$ invariant under the coordinate shift. Under the polynomial correspondence such a code can be regarded as an ideal in the ring $\mathbb{F}_{p}[x]/(x^N-1).$ It can be shown that this ideal is principal, with a unique monic generator $g(x),$ called {\bf the generator polynomial} of the code. The {\bf check polynomial} $h(x)$ is then defined as the
 quotient $(X^N-1)/g(x).$ A well-known fact is that the codewords are the periods of the linear recurrence of characteristic polynomial the reciprocal polynomial $h(x)$ \cite[p. 195]{MS}.
 A cyclic code is {\bf irreducible} if its check polynomial $h(x)$ is. A code is two-weight if it has only two non-zero weights. In \cite{SW} a conjectural classification scheme of irreducible cyclic two-weight codes is given as
 \begin{enumerate}
 \item a list of eleven exceptional codes
 \item subfield codes
 \item semiprimitive irreducible codes
 \end{enumerate}
 {\bf Subfield codes} is the case when the root of $h(x)$ is a primitive root of a subfield of the quotient field $\mathbb{F}_{p}[x]/(h(x)).$
To define semiprimitive codes write $p^m-1=N u,$ where $m$ is the dimension of the code. If $-1$ is a power of $p$ modulo $u,$ then the irreducible cyclic code of parameters $[N,m]$ is said to be {\bf semiprimitive}. In this paper, we will exhibit several codes that do not fit this classification.
\section{Main results}
 Let $p$ be a prime $>5.$ Let \{{$g_{n}$}\} denote the generalised Fibonacci numbers $G_n$ reduced modulo $p.$ This sequence is periodic of period a divisor of $\pi(p).$ Consider the cyclic code $C_p$ of parity check polynomial $h(x)=x^{2}+x-1$ of length $\pi(p)$ over $\mathbb{F}_{p}$. Thus the words of this code are the periods of $g_n$ when initial conditions vary. Similarly, let $f_n$ denote the Fibonacci numbers $F_n$ reduced modulo $p.$

 If $p=\pm 1 \pmod{10}$, the polynomial $h$ has two distinct roots in  $\mathbb{F}_{p}$; if $p=\pm 3 \pmod{10}$, it is irreducible over $\mathbb{F}_{p}$,
with two distinct roots over $\mathbb{F}_{p^2}.$ This alternative comes from the discriminant of $h$ which is $5.$ This integer is a square in $\mathbb{F}_{p}$ if $p=\pm 1 \pmod{10},$ and a non-square if $p=\pm 3 \pmod{10}$
(see \cite[Lemma 3.9]{R} for the elementary proof based on quadratic reciprocity).
\subsection{Irreducible cyclic code}
 If $p=\pm 3 \pmod{10}$, we know the roots modulo $p$ of $x^{2}+x-1$ do not belong to $\mathbb{F}_{p}$, and belong to the finite field $\mathbb{F}_{p}[x]/(x^{2}+x-1)\sim \mathbb{F}_{p^2}$. Obviously the roots of $x^{2}+x-1$ over $\mathbb{F}_{p}[x]/(x^{2}+x-1)$ are different, since $p>5.$ Let $r,s$ be the roots of $x^{2}+x-1$ over $\mathbb{F}_{p}[x]/(x^{2}+x-1)$. We have $r+s=-1 ,\, rs=-1 ,\, ord(r)=ord(s)=\pi(p) .$ From the finite field analogue of Binet's formula we know that $f_{n}=\frac{r^{n}-s^{n}}{r-s}.$
Let $N=\pi(p),\, e=ord(rs^{-1}).$ Observe that $e$ is a divisor of $N$.

\hspace{-0.53cm}\textbf{Lemma 1.1} If $f_{n}=0,$ then $n$ is a multiple of $e.$

\begin{proof}

From $f_{n}=\frac{r^{n}-s^{n}}{r-s}$ we see that  $f_{n}=0\Leftrightarrow (rs^{-1})^{n}=1$. Hence $n$ is a multiple of $e.$
\end{proof}

\hspace{-0.52cm}\textbf{Lemma 1.3}
$G=\left(
                                         \begin{array}{ccccc}
                                           f_{0} & f_{1} & f_{2} & ... & f_{N-1}  \\
                                           f_{N-1}& f_{0} & f_{1} & ... & f_{N-2}  \\
                                         \end{array}
                                       \right)$ is a generator matrix of the code $C_p.$

\begin{proof}
  By the cyclicity of the code $C_p$ the matrix $G$ spans a subcode of $C_p.$ It is of rank 2, since \[
\left|\begin{array}{cccc}
    f_{1} &    f_{2}   \\
    f_{0} & f_{1}
\end{array}\right|=1\neq 0.
\]  So $rank(G)=dim(C)=2.$ Thus $G$ is the generator matrix of the code $C_p$.
\end{proof}

\hspace{-0.53cm}\textbf{Theorem 1.4} If $p=\pm 3 \pmod{10},$  then $C_p$ is either a one-weight code with weight $\{N-K\}$; or a two weight code with weights $\{N-K,N\}.$

\begin{proof}
 Consider an element $\overrightarrow{z}$ of $C$. From Lemma 1.3, we know that $$\overrightarrow{z}=\lambda_{1}
(f_{0},f_{1},\cdots,f_{N-1})+\lambda_{2}(f_{N-1},f_{0}, f_{1},\cdots,,f_{N-2}),\lambda_{1},\lambda_{2}\in \mathbb{F}_{p}.$$
From Lemma 1.1, we know that $f_{0}=f_{e}=f_{2e}=\cdots=f_{(K-1)e}=0$, and that $f_{ie+j}\neq 0,0\leq i\leq K-1,1\leq j\leq e-1.$
Denote by $w(.)$ the Hamming weight of an element of $\mathbb{F}_{p}^N.$
Consider three cases:

(1) If $\lambda_{1}=\lambda_{2}=0$, then $\overrightarrow{z}=\overrightarrow{0},w(\overrightarrow{z})=0$.

(2) If $\lambda_{1}=0,\lambda_{2}\neq0~or~\lambda_{1}\neq0,\lambda_{2}=0$, then by the above discussion, $w(\overrightarrow{z})=N-K$.

(3) If $\lambda_{1}\neq0,\lambda_{2}\neq0$, then write $g_n$ for the sequence attached to the codeword $\overrightarrow{z}.$ Computing indices modulo $N,$ we see that $g_n=0$ iff
\[\lambda_{2} \frac{(r^{n}-s^{n})}{r-s}+\lambda_{1}\frac{(r^{n+1}-s^{n+1})}{r-s}=0, \]
or, equivalently, iff
\[(\lambda_{2}+\lambda_{1}r)r^n-(\lambda_{2}+\lambda_{1}s)s^n=0.\]
Since $r\in \mathbb{F}_{p^2}\setminus \mathbb{F}_{p},$ we see that $(\lambda_{2}+\lambda_{1}r)\neq 0.$ Thus
$g_n=0$ iff \[ \big(\frac{r}{s}\big)^n=\frac{\lambda_{2}+\lambda_{1}s}{\lambda_{2}+\lambda_{1}r}=0.\]
This equation in $n$ has either $K$ or no solutions in the integer range $\{0,\cdots,N-1\}.$ In the first case $w(\overrightarrow{z})=N-K;$ in the second case $w(\overrightarrow{z})=N.$
\end{proof}

\hspace{-0.53cm}{\bf Examples:} When $p=23$ and $n=48,$ we obtain a one-weight code. Same thing with $p=43,$ and $n=88.$
\subsection{Reducible cyclic code}
 If $p=\pm 1 \pmod{10}$, then the roots of $x^{2}+x-1$ belong to $\mathbb{F}_{p}$. And these roots are different since $p>5.$ Let $r,s$ be the roots of $x^{2}+x-1$ over $\mathbb{F}_{p}$, then we have $r+s=-1,rs=-1$ and $ord(r)=ord(s)=\pi(p)$.
Let $N=\pi(p), e=ord(rs^{-1})$, then $e$ is a divisor of $N$.

\hspace{-0.53cm}\textbf{Lemma 2.1} The matrix $G=\left(
                                         \begin{array}{ccccc}
                                           r & r^{2} & ... & r^{N-1} & r^{N} \\
                                           s & s^{2} & ... & s^{N-1} & s^{N} \\
                                         \end{array}
                                       \right)$ is a generator matrix of the code $C_p.$

\begin{proof} From the finite field analogue of Binet's formula we know that $g_{n}=ar^{n}+bs^{n},a,b\in \mathbb{F}_{p}$. Thus the matrix $\left(
                                         \begin{array}{ccccc}
                                           r & r^{2} & ... & r^{N-1} & r^{N} \\
                                           s & s^{2} & ... & s^{N-1} & s^{N} \\
                                         \end{array}
                                       \right)$ spans a subcode of $C_p.$
It is of rank 2 since \[
\left|\begin{array}{cc}
    r &    r^{2}   \\
    s & s^{2}
\end{array}\right|=-rs(r-s)\neq 0.
\]So rank$(G)=\dim(C)=2.$ Thus $G$ is a generator matrix of the code $C_p$.
\end{proof}

\hspace{-0.53cm}\textbf{Theorem 2.2} The code $C_p$ is a two-weight code with weights $\{N-K,N\}$.

\begin{proof}
Consider an element $\overrightarrow{z}$ of $C_p.$ From Lemma 2.1 we know that $$\overrightarrow{z}=\lambda_{1}
(r,r^{2},...,r^{N})+\lambda_{2}(s,s^{2},...,s^{N}),\lambda_{1},\lambda_{2}\in \mathbb{F}_{p}.$$

Thus $g_n=\lambda_1 r^n+\lambda_2 s^n.$
If only one of $\lambda_{1},$ or $\lambda_{2}$ is zero then $g_n$ is never zero and $w(\overrightarrow{z})=N.$
Assume $\lambda_{1}\lambda_{2}\neq 0.$ Then $g_n=0$ iff \[(\frac{r}{s})^n=-\frac{\lambda_{2}}{\lambda_{1}}.\]
This equation in $n$ admits either zero or $N-K$ solutions.
Altogether, $C_p$ is a two weight code with weights $\{N-K,N\}$.
\end{proof}

\subsection{General properties of the code $C_p$}
We study the dual distance of $C_p.$

\hspace{-0.53cm}\textbf{Theorem 3.1} The dual distance of $C_p$ is at least $2$ and at most three. It is three when $K=1.$ In that case $C_p$ is MDS, and $C_p^\bot$ is uniformly packed.
\begin{proof}
In both congruence cases of $p$ it can be checked that the dual distance of $C_p$ is at least $2,$ by exhibiting nonsingular minors of order $2$ in $G$ like in the proof of Lemmas 1.3 and 2.1.
It is at most three by the Singleton bound applied to $C_p^\bot,$ with equality iff $C_p$ is MDS \cite[Chap 11, Th. 2]{MS}. Since the minimum distance of $C_p$ is $N-K$ this happens iff $K=1.$
The dual of a two-weight code with minimum distance three is uniformly packed \cite{Cald}.
\end{proof}

\hspace{-0.53cm}{\bf Remark:} If $p \equiv \pm 3 \pmod{10},$ the code $C_p$ is never MDS. In this case it can be shown that $\pi(p)$ is a multiple of $4$ \cite{R}, and this implies that $\alpha(p)$ divides $N/2,$ because
$ord(r/s)=ord(-r^2),$ by $rs=-1.$

We compute the weight distribution when $C_p$ is a two-weight code.

\hspace{-0.53cm}\textbf{Theorem 3.2} If the code $C_p$ is a two-weight code with weights $\{N-K,N\}$, then the respective frequencies are \{$\frac{(p-1)N}{K}, \frac{(p-1)(K(p+1)-N}{K}$\}.
\begin{proof} By Theorem 3.1 the dual distance of $C_p$ is at least $2.$
 The result follows then by application of the Pless power moments and resolution of the system in the frequencies $x,y$ given by
\begin{eqnarray*}
x+y&=&p^2-1\\
x(N-K)+yN&=&p(p-1)N.
\end{eqnarray*}
This proves the results. \end{proof}

We show that the coset graph of $C_p^\bot$ is a strongly regular graph, and we compute its non trivial eigenvalues.

\hspace{-0.53cm}\textbf{Theorem 3.3} If $K=1,$ the coset graph of $C_p^\bot$ is a simple strongly regular graph on $p^2$ vertices of degree $N(p-1)$ with nontrivial eigenvalues $\{ -N, p-N\}.$

\begin{proof} By Theorem 3.1, the code $C_p$ is projective. The result follows by
\cite[Cor. 9.8.2]{BH}, and the values of the weights $\{N-K, K\}.$
\end{proof}

\section{Numerical examples}
\subsection{Irreducible cyclic codes}
To appreciate the following examples we make the following remarks in the case $p \equiv \pm 3 \pmod{10}.$
\begin{enumerate}
\item By definition of $u$ we have $Nu=p^2-1=2(p+1)(\frac{p-1}{2}).$ This implies, in particular, that if $N=2(p+1),$ we have $p=2u+1,$ hence $p \equiv 1 \pmod{u}.$ Thus, such an irreducible code cannot be semiprimitive.
 \item   Further none of our Pisano codes can be subfield codes in the sense of \cite{SW}, since it is known \cite{R}, that if $p \equiv \pm 3 \pmod{10},$ the Pisano period  $N=\pi(p)$ divides $2(p+1)$, which is coprime with   $p-1.$ \item  Eventually, all the eleven exceptional codes in \cite{SW} have dimension $>2.$
\end{enumerate}

 The codes in Tables 1 and 2 below all satisfy $N=2(p+1),$ except for $p=47,107,113.$ Thus we claim to have found at least 17 counterexamples to the Schmidt-White scheme.

\begin{center}
\small  {Table $1$} \\ \vspace{0.2cm}
\begin{tabular}{|c|c|c|c|c|c|c|c|c|c|c|}
  \hline
   $p$                & 3 & 7  & 23 & 43 & 67  & 83  & 103 & 127 & 163 & 167 \\
  \hline $N$          & 8 & 16 & 48 & 88 & 136 & 168 & 208 & 256 & 328 & 336 \\
  \hline $e$          & 4 & 8  & 24 & 44 & 68  & 84  & 104 & 128 & 164 & 168 \\
  \hline
\end{tabular}
\end{center}
\vspace{0.2cm}

\begin{center}
\small \small {Table $2$} \\ \vspace{0.2cm}
\begin{tabular}{|c|c|c|c|c|c|c|c|c|c|c|}
  \hline
   $p $               & 13 & 17  & 37 & 47 & 53  & 73  & 97 & 107 & 113 & 137 \\
  \hline $N $         & 28 & 36 & 76 & 32 & 108 & 148 & 196 & 72 & 76 & 276 \\
  \hline $e $         & 7 & 9  & 19 & 16 & 27  & 37  & 49 & 36 & 19 & 69 \\
  \hline
\end{tabular}
\end{center}

\subsection{Reducible cyclic codes}

Tables 3 and 4 show that the code $C_p$ is MDS ($d^{\bot}=3$) for many primes $p \equiv \pm 1 \pmod{10}.$
\begin{center}
\small   \small {Table $3$} \\ \vspace{0.2cm}
\begin{tabular}{|c|c|c|c|c|c|c|c|c|c|c|}
  \hline
   $p$          & 11 & 19 & 29 & 31 & 41 & 59 & 61 & 71 & 79 & 89 \\
  \hline $N$          & 10 & 18 & 14 & 30 & 40 & 58 & 60 & 70 & 70 & 44 \\
  \hline $e $         & 10 & 18 & 14 & 30 & 20 & 58 & 15 & 70 & 70 & 11 \\
  \hline $d^{\bot}$ & 3  & 3  & 3  & 3  & 2  & 3  & 2  & 3  & 3  & 2  \\
  \hline
\end{tabular}
\end{center}
\vspace{0.2cm}

\begin{center}
 \small {Table $4$} \\ \vspace{0.2cm}
\begin{tabular}{|c|c|c|c|}
\hline  $p$ & $N$ & $e$ & $d^{\bot}$  \\
\hline\hline
      11 & 10& 10&3   \\
\hline19 & 18& 18&3   \\
\hline29 & 14& 14&3   \\
\hline31 & 30& 30&3   \\
\hline41 & 40& 20&2   \\
\hline59 & 58& 58&3   \\
\hline 61& 60& 15&2   \\
\hline 71& 70& 70&3   \\
\hline 79& 78& 78&3   \\
\hline 89& 44& 11&2   \\
\hline
\end{tabular}
\end{center}

\section{Conclusion and open problems}

In this paper we have constructed an infinite family of irreducible two-weight codes over prime fields. In particular, we have
exhibited a number of counterexamples to the Schmidt-White conjecture on irreducible cyclic codes with two weights. To obtain an infinite number of counterexamples we pose the following number theoretic question, of interest in its own right.

{\bf Question:} Are there infinitely many primes $p\equiv \pm 3 \pmod{10}$ such that $\pi(p)=2(p+1)?$

As observed before, for such a prime $p,$ the code $C_p$ is an irreducible cyclic code that is neither subfield nor semiprimitive. A positive answer to that question would provide an infinity of counterexamples to the Schmidt-White classification.  More generally, the methods we use are new in coding theory and very old in number theory. It would be interesting to know to which class of cyclic codes the concept of rank can be extended.

\end{document}